\documentclass[a4paper]{article}
\usepackage{footmisc}
\usepackage{graphicx}

\begin{document}
\title{The MHD nature of ionospheric wave packets excited by the solar terminator}

\author{E.L. Afraimovich, S.V. Voyeikov, I.K. Edemskiy, Yu.V. Yasyukevich\\ \\
Institute of Solar-Terrestrial Physics SB RAS, Irkutsk, Russia,\\
p.o.~box~291, 664033, fax: +7 3952 511675.\\ e-mail:~ilya@iszf.irk.ru}

\maketitle

\section*{Abstract}
We obtained the first experimental evidence for the
magnetohydrodynamic (MHD) nature of ionospheric medium-scale
travelling wave packets (MSTWP). We used data on total electron
content (TEC) measurements obtained at the dense Japanese network
GPS/GEONET (1220 stations) in 2008-2009. We found that the
diurnal, seasonal and spectral MSTWP characteristics are specified
by the solar terminator (ST) dynamics. MSTWPs are the chains of
narrow-band TEC oscillations with single packet's duration of
about 1-2 hours and oscillation periods of 10-20 minutes. Their
total duration is about 4--6 hours. The MSTWP spatial structure is
characterized by a high degree of anisotropy and coherence at
the distance of more than 10 wavelengths. The MSTWP direction of
travelling is characterized by a high directivity regardless of
seasons. Occurrence rate of daytime MSTWPs is high in
winter and during equinoxes. Most of daytime MSWPs propagate
southeastward (155$\pm$28${}^\circ$) at the velocity of 130$\pm$52 m/s.
Occurrence rate of nighttime MSTIDs has its peak in summer. They
propagate southwestward (245$\pm$15${}^\circ$) at the velocity of
110$\pm$43 m/s. These features are consistent with previous MS
travelling ionosphere disturbance (TID) statistics obtained from
630-nm airglow imaging observations in Japan. In winter, MSTWPs in
the northern hemisphere are observed 3-4 hours after the morning
ST passage. In summer, MSTWPs are detected 1.5-2 hours before the
evening ST occurrence at the point of observations, at the
moment of the evening ST passage in the magneto-conjugate point.
Both the high Q-factor of oscillatory system and synchronization
of MSTWP occurrence with the solar terminator passage at the point of
observations and in the magneto-conjugate area testify the MHD
nature of ST-excited MSTWP generation. The obtained results are
the first experimental evidence for the hypothesis of the
ST-generated ion sound waves.

\section{Introduction}
\label{SPE-sect-1}

    Pioneering theoretical works of V.M. Somsikov ((1995) and other
papers) marked the beginning of numerous experimental observations
of "terminator" waves with the use of different methods of
ionospheric radio sounding. Recent investigations have shown that
movement of the solar terminator (ST) causes generation of
acoustic-gravity waves (AGW), turbulence and instabilities in the
ionosphere plasma. It is worth noting that among all the sources
of gravity waves, the moving ST has a special status, since it is
a predictable phenomenon whose characteristics are well known.
Considering the ST as a stable and repetitive source of AGWs, one
can derive information about atmospheric conditions from the
response of the medium to this input. The great variety of
ST-linked phenomena in the atmosphere gave rise to a number of
studies on the analysis of ionosphere parameter variations
obtained by different methods of ionosphere sounding [Drobzhev et
al., 1992; Hocke and Schlegel, 1996; Dominici et al., 1997;
Galushko et al., 1998]. However, virtually all experimental data
were obtained using indirect methods for analyzing the spectrum of
ionosphere parameter variations, which can result from a number of
factors. This causes difficulties in the reliable identification
of ST AGWs, because in general case AGWs can be generated by
different sources either of natural or of anthropogenic origin
[Hocke and Schlegel, 1996].

Recently a considerable progress has been achieved in the
study of ionosphere irregularities using the new technology of GPS
radio sounding. It allows us to obtain the data on variations of
the total electron content (TEC) with high spatial and temporal
resolution. Afraimovich (2008) has obtained the first GPS TEC
evidence for the wave structure excited by the morning ST, moving
over the USA, Europe, and Japan. The author has first
found ST-generated medium-scale travelling wave packets (MSTWP).
These MSTWPs have duration of about 1-2 hours and time shift of
about 1.5-2.5 hours after the morning ST appearance at the altitude of 100
km.

Registering time dependence of ionosphere parameters is insufficient
to identify ST-generated wave disturbances. It is
necessary to determine the spatial structure of these disturbances
and compare it with the spatial-temporal characteristics of ST.
Hence, it is very important to define the spatial structure of MS
wave packets in TEC. Using TEC measurements from the dense network
of GPS sites GEONET, Afraimovich et al. (2009) has obtained the
first GPS-TEC image of the space structure of MSTWP excited by the
morning ST motion of ST on 13 June 2008.

The goal of this paper is to obtain the detailed information
regarding the spatial, spectral and dynamic characteristics of
MSTWPs excited by the ST as deduced from the dense GPS network GEONET.

\section{Data and Methods}
\label{SPE-sect-2}

We use data from the Japanese GPS network GEONET (about 1225
stations in total). Actually, it is the world's largest regional GPS
network\footnote[2]{\texttt{ftp://terras.gsi.go.jp/data/GPS\symbol{`\_}products/}}. The
geomagnetic situation on selected and analyzed days of 2008-2009
can be characterized as quiet: the Kp index varied from 1.0 to
3.0.

The standard GPS technology provides a means of detection of wave disturbances
based on phase measurements of TEC at each of
two-frequency spaced GPS receivers [Calais et al., 2003; Afraimovich et
al., 2003; Hernandez-Pajares et al., 2006]:

\begin{equation}
\label{IID-eq-01}
             I_o=\frac{1}{40{.}308}\frac{f^2_1f^2_2}{f^2_1-f^2_2}
                           [(L_1\lambda_1-L_2\lambda_2)+const+nL],
\end{equation}
where $L_1\lambda_1$ and $L_2\lambda_2$ are additional paths of
the radio signal caused by the phase delay in the ionosphere,~(m);
$L_1$ and $L_2$ represent the number of phase rotations at the
frequencies $f_1$ and $f_2$; $\lambda_1$ and $\lambda_2$ stand for
the corresponding wavelengths,~(m); $const$ is the unknown initial
phase ambiguity,~(m); and $nL$~ are errors in determining the
phase path,~(m). Phase measurements in GPS can be made with a
high degree of accuracy corresponding to the error of TEC
determination of at least $10^{14}$~m${}^{-2}$, when averaged on a
30-second time interval, but with some uncertainty of the initial
value of TEC. This allows us to detect ionization
irregularities and wave processes in the ionosphere over a wide
range of amplitudes (up to $10^{-4}$ of the diurnal TEC variation)
and periods (from 24 hours to 5 min). The total electron content
unit (TECU) equal to $10^{16}$~m${}^{-2}$~ and commonly
accepted in the literature will be used below.

Primary data include series of "oblique" values of TEC
$I_0(t)$, as well as the corresponding series of elevations
$\theta_s(t)$ and azimuths $\alpha_s(t)$ of the line of sight
(LOS) to the satellite.

Series of values of elevations $\theta_s(t)$ and azimuths
$\alpha_s(t)$ of LOS to the satellite was used to determine
coordinates of subionospheric points (SIP) for the height
$h_{max}$=300 km of the $F_{2}$-layer maximum and to convert the
"oblique" TEC $I_{o}(t)$ to the corresponding value of the
"vertical" TEC

\begin{equation}
\label{IID-eq-02}
I = I_o \times cos
\left[arcsin\left(\frac{R_z}{R_z + h_{max}}cos\theta_s\right)
\right],
\end{equation}
where $R_{z}$ is the Earth's radius. All results in this study
were obtained for elevations $\theta_s(t)$ larger than 50$^\circ$.

To eliminate variations in the regular ionosphere, as well as
trends introduced by orbital motion of the satellite, we derive
TEC variations $dI(t)$ by filtering from the initial
$I(t)$-series over the range of periods of 2-20 min.

\begin{figure}[!t]
\includegraphics[clip, width=\textwidth]{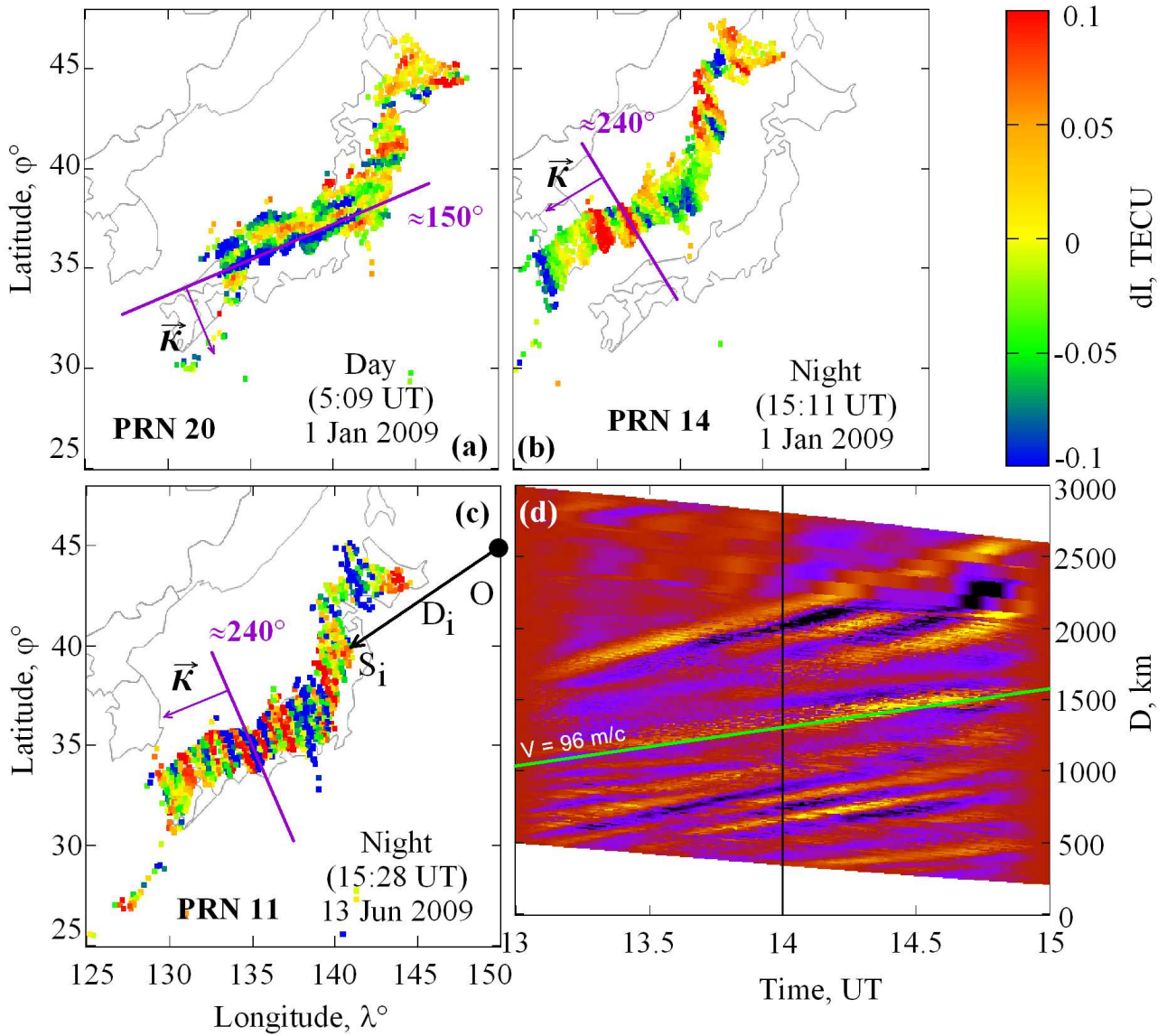}
\caption {The latitude-longitude 2D-space distribution of the
filtered TEC $dI$ ($\varphi$, $\lambda$) after morning ST passage
over Japan on 1 January 2009, 05:09 UT ({\bf a}); after evening ST
on 1 January 2009, 15:11 UT ({\bf b}); and after evening ST on 13
June 2008, 15:28 UT ({\bf c}). ({\bf d}) - TWPs for all GPS sites
located along the direction $D$ over Japan for selected satellite
number PRN19 on 13 June 2008, 13:00-15:00 UT. The velocity $V$ of
TWPs is defined through the inclination of the wave front line
(the green line on the panel ({\bf d})).}
\end{figure}

\section{TWP spatial characteristics}
\label{SPE-sect-3}

We use two known forms of the 2D GPS-TEC imaging for our
presentation of the ST-generated MSTWP space structure. First we
apply the diagram "distance-time" $dI(t, D)$ used by
different authors [Calais et al., 2003; Afraimovich et al., 2009].
This diagram is calculated in the polar coordinate system (Fig.
1c). Points $0$ and $S_i$ mark the relative coordinate centre and
the SIP position for LOS between the GPS site with number $i$
and the selected satellite PRN, respectively; $D_i$ is the
distance between points $0$ and $S_i$ along the great circle arc.
Here we select the geographic latitude and longitude for point
$0$ (45$^\circ$N and 150$^\circ$E) outside the GEONET GPS sites
field. This location of the relative coordinate center is chosen
to make more adequate correspondence between distance to each GPS
site and its actual location.

Neglecting the Earth's sphericity, we can determine
coordinates $x_i(t)$ and $y_i(t)$ of the SIPs $S_i$ at an altitude
$h_{max}$ in the topocentric coordinate system with the center in
GPS as

\begin{equation}
\begin{array}{rl}
x_i(t) = h_{max} sin(\alpha_i(t))cot(\theta_i(t)) \\
y_i(t) = h_{max} cos(\alpha_i(t))cot(\theta_i(t))
\end{array}
\label{IID-eq-03}
\end{equation}
where $\alpha_s$ is the azimuth counted off from the north in a
clockwise direction; $\theta_s$ is the LOS elevation
[Afraimovich et al., 2003].

Then we recalculate topocentric coordinates $x_i(t)$ and
$y_i(t)$ into corresponding values of the latitude $\varphi$ and
longitude $\lambda$ of SIPs $S_i$ , calculate distance
$D_i$ and then plot the 2D-diagram "distance-time" $dI(t, D)$ for
the selected satellite number PRN. Our approximation is acceptable
only for large $\theta_i$ values. So for the minimum
$\theta_i$ value of about 45$^\circ$ and $h_{max}$ = 300 km, the
maximum deviation of SIPs $S_i$ from GPS site does not exceed
300 km.

In order to study space properties and dynamics of TWP packets
in detail, we also employ another form of the 2D-space distribution
of values of filtered TEC series $dI$ ($\varphi$, $\lambda$)
in latitude $\varphi$ and longitude $\lambda$ for each 30-sec
TEC counts with spatial resolution of 0.15$^\circ$ in latitude and
0.15$^\circ$ in longitude. Then we build up the dynamic image of
disturbances travelling over Japan in video format ($AVI$).
Similar technology was first developed by Saito et al.(1998).
Authors showed that a high-resolution two-dimensional mapping of TEC
perturbations at the GPS GEONET network revealed spatial and
temporal TEC variations at mid-latitudes in detail. This had never
been attained in the past. The said space-time presentation
method of ionosphere disturbances for the dense GPS networks of USA was
used in [Tsugawa et al., 2007].

Fig. 1 presents the latitude-longitude 2D-space distribution
of the filtered TEC $dI$ ($\varphi$, $\lambda$) after the morning ST
passage over Japan on 1 January 2009, at 05:09 UT ({\bf a}); after
the evening ST on 1 January 2009, at 15:11 UT ({\bf b}); and after
the evening ST on 13 June 2008, at 15:28 UT ({\bf c}). The amplitude of
$dI(t)$ variations is about 0.1 TECU (scale for $dI$ is plotted
on the panel ({\bf b}) at the right). Lines mark the wave front
extension; numbers near lines mark direction $\alpha$ of the MSTWP
wave vector {\boldmath $K$}: 150$^\circ$ for 1 January 2009 (05:09
UT), 240$^\circ$ for 1 January 2009 (15:11 UT) and 240$^\circ$ for
13 June 2008 (15:28 UT), respectively.

Fig.1({\bf d}) presents the 2D-diagram "distance-time" $dI(t,
D)$ for TWPs on 13 June 2008, for GPS sites located along the
direction $D$ over Japan for satellite number PRN19
at 13:00-15:00 UT. The velocity $V$=96 m/s of TWPs is defined
through the inclination of the wave front line (the green line on
the panel ({\bf d})). Vertical line marks the time moment
15:28 UT (see the corresponding 2D-space distribution $dI$ ($\varphi$,
$\lambda$) on the panel ({\bf c})).

We have analyzed a number of similar images of MS TWP spatial
structure over Japan. The MS TWP space image is
characterized by pronounced anisotropy (ratio between the
lengthwise and transversal scales exceeds 10) and high coherence
over a long distance of about 2000 km. The MS TWP wavelength varies
from 100 km to 300 km. Spatial structure of 2D-space distribution
$dI$ ($\varphi$, $\lambda$) remain stabpe along whole Japan
at the distance of up to 2000 km. 

\section{Dynamic characteristics of TWPs by the
SADM-GPS method} \label{SPE-sect-4}

In order to determine velocity of TIDs, we use method taking
into account the relative motion of SIPs into account [Afraimovich et al., 1998,
2003]. We determine velocity and direction of the
wave front motion in terms of some model, whose adequate choice is
of crucial importance. In the simplest form, space-time
variations in TEC variations $dI(t, x, y)$ at each given moment of time $t$
can be represented in terms of the phase pattern that moves
without any change in its shape (non-dispersive disturbances):

\begin{equation}
I(t,x,y)=F(t-x/u_x-y/u_y) \label{IID-eq-04}
\end{equation}
where $u_x(t)$ and $u_y(t)$ are displacement velocities of
intersection of the phase front of axes $x$ and $y$,
respectively; $F$ is an arbitrary function.

A special case of (4) is the most often used model for a solitary
plane travelling wave of the TEC disturbance

\begin{equation}
I(t,x,y)=\delta\sin(\Omega t-K_xx-K_yy+\varphi_0)
\label{IID-eq-05}
\end{equation}
where $\delta$, $K_x$, $K_y$, $\Omega$ are the
amplitude, the $x$- and $y$- projections of the wave vector
{\boldmath $K$}, and the angular frequency of the disturbance, respectively;
$\varphi_0$ is the initial disturbance phase.

A SADM-GPS method was proposed by Afraimovich et al. (2003) for
determining the TID dynamics in the horizontal plane by measuring
variations of TEC derivatives with respect to the spatial
coordinates $I'_x(t)$, $I'_y(t)$ and to the time $I'_t(t)$. This
allows us to determinate the unambiguous
$\alpha(t)$ orientation of the wave-vector {\boldmath $K$} over the range
of 0--360${}^\circ$ and the horizontal velocity $V(t)$ at each
specific instant of time. A detailed description of the method is
presented in [Afraimovich et al. 2003].

Let us take a brief look at the sequence of data handling
procedures. Out of a large number of GPS stations, three points
($A, B, C$) are chosen in such a way that the distances between
them do not exceed about a half of the expected wavelength
$\Lambda$ of the disturbance. The point $B$ is considered as the
center of the topocentric frame of reference. Such configuration
of GPS receivers represents a GPS-array (or a
GPS-interferometer) with the minimum of the necessary number of
elements. In regions with a dense network of GPS-points, we can
obtain a broad range of GPS-arrays of a different configuration,
which provides a means for testing the data obtained for
reliability; in this paper we took advantage of this
possibility.

The input data include series of vertical TEC $I_A(t)$,
$I_B(t)$, $I_C(t)$. Linear transformations of the differences between
values of filtered TEC variations $(dI_{{\rm B}}-dI_{{\rm
A}})$ and $(dI_{{\rm B}}-dI_{{\rm C}})$ at the receiving points A,
B and C are used to calculate the components of the TEC gradients
$I'_x$ and $I'_y$.

The resulting series are used to calculate instantaneous values of
the horizontal velocity $V(t)$ and the azimuth $\alpha(t)$ of TID
propagation. Next, the series $V(t)$ and $\alpha(t)$ are put to a
statistical treatment. This involves construction of distributions of
the horizontal velocity $P(V)$ and direction $P(\alpha)$, which are
analyzed to test the hypothesis of the existence of the preferred
propagation direction. If such a direction does exist, the
corresponding distributions are used to calculate the mean values
of the horizontal velocity $\langle V\rangle$ and azimuth $\langle \alpha\rangle$ of TID propagation.
Using different sets of $n$ GPS arrays, we managed to obtain
average values of the horizontal projection $V$ and direction
$\alpha$.

\begin{figure}[!t]
\includegraphics[clip, width=0.75\textwidth]{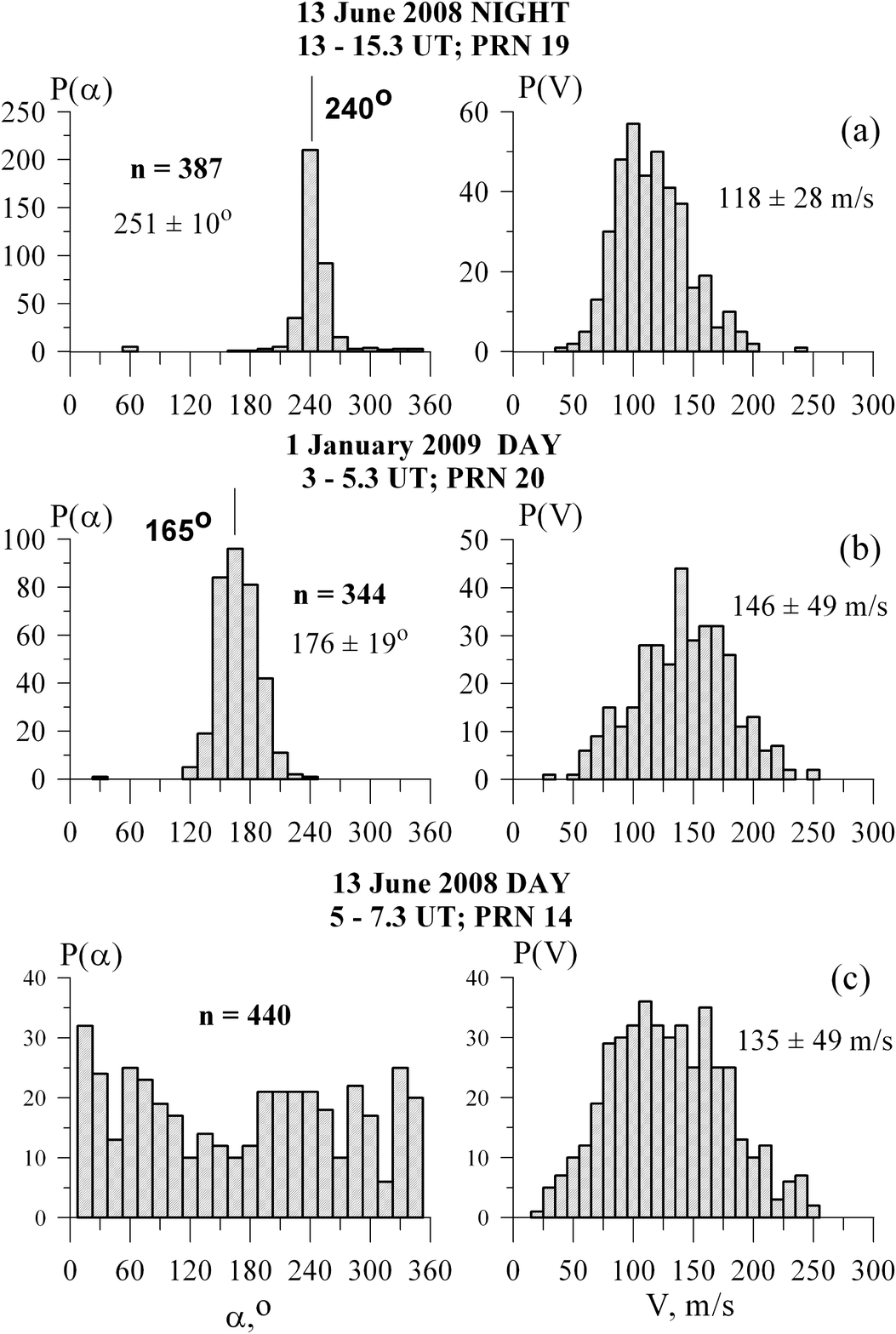}
\centering
\caption {Distributions of the TWP parameters for the nighttime of
13 June 2008 ({\bf a}), for the daytime of 1 January 2009 ({\bf b}),
and for the daytime of 13 June 2008 ({\bf c}), as determined by the
SADM-GPS method; azimuth $\alpha$ - at the left, and horizontal
component of TWP travelling velocity $V$ - at the right. Events
({\bf a}) and ({\bf b}) correspond to the strongly directional
travelling of TWPs at the night and the day; the event ({\bf b})
illustrates the absence of directional TWPs travelling.}
\end{figure}

Fig. 2. presents the examples of the distributions of the
MSTWP dynamics parameters for nighttime of 13 June 2008, 13-15.3
UT ({\bf a}), $n$=387; for daytime of 1 January 2009, 03-05.3 UT
({\bf b}), $n$=344; and for daytime of 13 June 2008, 05-7.3 UT
({\bf c}), $n$=440; azimuth $\alpha$ - at the left, and horizontal
component of MSTWP travelling velocity $V$ - at the right. Events
({\bf a}) and ({\bf b}) correspond to the strongly directional
travelling of MSTWPs for nights and days; the mean value and
RMS of the azimuth $\alpha$ equal 240$^\circ$ and 10$^\circ$, and
165$^\circ$ and 19$^\circ$, respectively. The event ({\bf b})
illustrates a total absence of directional TWPs travelling. In
all events, the velocity $V$ does not exceed 250-300 m/s. The mean
value and RMS of $V$ equal 118 and 28 m/s for the night; 146 and
49 m/s for the day, and 135 and 49 m/s for nondirectional
travelling, respectively.

The hypothesis of the connection between wave packet generation
and ST appearance can be tested in the terminator local time
($TLT$) system: $dT=t_{obs}-t_{st}$, where $t_{obs}$ is the data
point time, and $t_{st}$  is the time of ST appearance at the
altitude of $H$ over this point. In other words, firstly we
transformate the latitude and longitude of the point to a time of
the terminator appearance over this point and then we define the
difference between the terminator appearance time and the time at the
data point. A distinctive feature of this approach is in excluding
the point coordinates and considering each point data in the
solar terminator context only [Afraimovich et al., 2009]. Of the
greatest interest is total diurnal distributions of
travelling direction $\alpha$ for summer 2008 (Fig. 3. ({\bf
a})) and winter 2009 (Fig. 3. ({\bf e})) versus the local
terminator time $dT$. 2D- distributions of the azimuth $\alpha$
are calculated using data from n=492590 GPS-arrays for ({\bf a}); and $n$=197150 for
({\bf e}). The numbers on the panels mark the numbers of the analyzed
days for the summer and winter. The steps on TLT $dT$ and
azimuth $\alpha$ equal 0.5 h and 5 $^\circ$, respectively. The
scale for number of count hits $N$ in different bins of
distributions is plotted on panels ({\bf a}) and ({\bf e}) at the
right.

The total distribution in Fig.({\bf a}) corresponds to the strongly
directional travelling of MSTWPs at the night for full 35 summer
days; the mean value and RMS of the azimuth $\alpha$ equal
240$^\circ$ and 10$^\circ$, respectively. The strongly directional
travelling begins immediately after the sunset (SS) ST appearance at
observational points and continues for about 5 h. A similar
picture is also significant for the night for full 16 winter
days (Fig.({\bf e})); the mean value and RMS of the azimuth $\alpha$
equal 240$^\circ$ and 10$^\circ$, respectively. But in this case
the strongly directional travelling begins 5 h after SS ST
appearance at observation points and continues for about 5
h. For the day, the strongly directional travelling of MSTWPs
corresponds to the mean value and RMS of the azimuth $\alpha$ equal to
160$^\circ$ and 10$^\circ$, respectively. It starts 4 h after
the sunrise (SR) ST appearance at observation points and continues
for about 3-4 h. Beyond the mentioned time intervals is a total absence of directional TWP travelling.

As we have found out, two types of MSTWPs - regular and chaotic
ones - are observed for 24 hours. The regular travel with
duration of some hours is synchronized with the ST passage. "Day"
and "night" values of the TWP wave vector direction in the regular
motion stage are securely fixed and independent of a season and
terminator front direction.

\begin{figure}[!t]
\includegraphics[clip, width=0.82\textwidth]{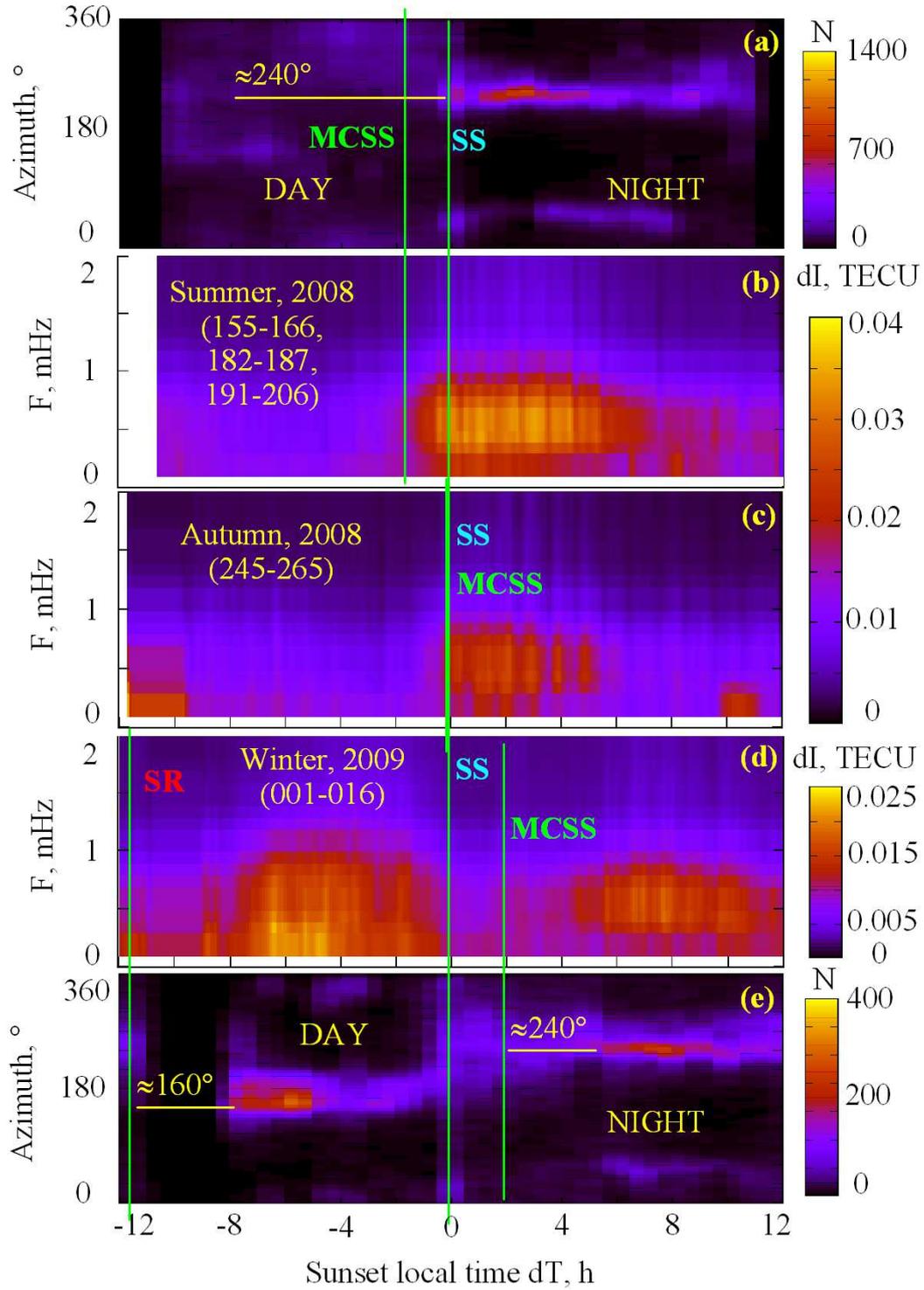}
\centering
\caption {({\bf b}), ({\bf c}), ({\bf d}) - the dynamic
spectra of TWPs versus the local terminator time $dT = t_{obs} -
t_{st}$ between the time of observation $t_{obs}$ and the time
$t_{st}$ of sunset (SS) ST passing over the observation points.
({\bf a}), ({\bf e}) - similar dependencies of the azimuth
$\alpha$ versus the local terminator time $dT$. Vertical lines
MCSS mark the moments of SS ST passage over the magneto-conjugate
point.}
\end{figure}

\section{TWP spectrum and ST appearance over observational and
magneto-conjugate points} \label{SPE-sect-5} Main
characteristics of wave processes are temporal and spatial
spectra. A calculation of a single spectrum of TEC variations
involves continuous series of $dI(t)$ with duration of no
less than 2.3 hours, thus enabling us to obtain a number of
counts equal to 256 that is convenient for the algorithm of the fast
Fourier transform (FFT) used in this study. Before spectral
analysis procedure, series $dI(t)$ are obtained by filtering from
the initial $I(t)$-series over the range of periods of 2-30 min.
To improve the statistical validity of the data, we use the
method involving a regional spatial averaging of TEC disturbance
spectra for all GPS/GEONET sites.

Figs. 3({\bf b}), ({\bf c}), ({\bf d}) present the dynamic spectra
of TWPs versus the local terminator time $dT$ for the sunset (SS) ST
passing over the point of observation and for SS over the magneto-conjugate point (MCSS). The dynamic spectra are calculated using
$N$ counts for each bins of spectra (total number of $dI(t)$ data
series $m$=3687173 for ({\bf b}); $m$=2059177 for ({\bf c});
$m$=1656426 for ({\bf d})). The numbers on the panels mark the
numbers of the analyzed days for the summer 2008 (({\bf b}); for the
autumn 2008 ({\bf c}), and for the winter 2009 (({\bf d}). The steps
on TLT $dT$ and variation frequency $F$ equal 0.5 h and 0.13 mHz,
respectively. The amplitude of spectral components is of about
0.02--0.1 TECU (scale for $dI$ is plotted at the right).

According to the dynamic spectra, MSTWPs are the chain of narrow-band
TEC oscillations with individual packet's duration of about 1-2
hours and oscillation periods of 10-20 minutes. Its total duration
is about 4--6 hours.

The most important discovery is that TWPs in Japan in summer are
registered 1.5-2 hours before the ST appearance over the
registration point, when the evening ST passes over the magneto-conjugate
 point in Australia ({\bf b}). At the equinox, those
moments (SS and MCSS) contemporize ({\bf c}); in winter TWPs
are recorded 4-6 hours after the evening SS and MCSS moments ({\bf
d}).

It is very interesting to compare $TLT$ dependencies of MSTWP
spectrum with similar dependencies of the azimuth $\alpha$ (Fig.
3({\bf a}), ({\bf e})). The strongly directional travelling begins
after MSTWP appearance and continues over less time interval
than the MSTWP appearance in spectra.

\section{Discussion}
\label{SPE-sect-6} Our observations confirm that the ST is a
stable and repetitive source of ionospheric wave disturbances.
Besides, we confirm the validity of the discovery of ST wave
packets made by Afraimovich (2008). The obtained results are in
agreement with the theoretical indications of solar terminator
effects [Somsikov, 1995] and do not contradict the results
obtained by Drobzhev et al. (1992) and Dominici et al. (1997),
which are based on limited statistical material. Drobzhev et al.
(1992) studied the dynamic spectra of the radio-wave reflection
virtual heights obtained by the vertical sounding of the ionosphere.
It was shown that during transient hours of the day under
magnetically quiet conditions, the low-frequency maximum of the
spectra shifted to the higher-frequency region.

Our main results agree with the data on the MSTID spatial structure,
dynamics and spectrum variations from different methods of
ionosphere radio sounding [Drobzhev et al., 1991, 1992;
Jacobson et al., 1995; Mercier, 1996; Hocke and Schlegel, 1996;
Dominici et al., 1997; Galushko et al., 1998; Saito et al., 1998;
Calais et al., 2003; Afraimovich et al., 1998, 1999, 2003;
Hernandez-Pajares et al., 2006; Kotake et al., 2007; Tsugawa et
al., 2007]. Kotake et al. (2007) first showned
statistical characteristics of the MSTIDs over Southern California
observed with densely spaced GPS receivers, and found that
characteristics of MSTIDs varied in daytime, dusk, and
nighttime.

On the other hand the recent advent of new observation techniques
using all-sky CCD imagers brought about new insights into the
two-dimensional properties of TIDs. In Japan there is a network
of 630.0 nm (OI) airglow imagers. Our results regarding the MSTWP
spatial structure and dynamics check well with the data from all-sky CCD
imagers [Ogawa et al., 2002; Otsuka et al., 2004; Shiokawa et al.,
2003, 2005].

The question about the mechanism of the observed TEC oscillation
is still unclear. MSTIDs (including those dependent on the
terminator) are traditionally connected with modulation of
the electron density of AGWs generated in the lower atmosphere as the
terminator passes over the observation point. However, this
hypothesis is not in agreement with characteristics of terminator
waves we have found out (a high spatial coherence, strong
anisotropy, stable directions of the wave vector azimuth). It is
known that AGWs and TIDs can propagate without significant
attenuation and changes in their shape or loss of their coherence
no farther than 3-5 wavelengths; MSTIDs can propagate within 500-1000 km
[Francis, 1974; Drobzhev et al., 1991].

The strongest argument against the AGW model of wave packets, at
least for night conditions in summer, is our observations of TWP
1.5 hour before the terminator goes over an observation point and
the explicit dependence on the instant of the terminator passage
through a magneto-conjugate point. In winter in the north
hemisphere, TWPs can be observed 3 hours after the morning ST
passage. In equinoxes, TWPs appear after the ST passage without a
considerable delay or advance. In summer, TWP can be recorded
1.5-2 hours before the evening ST appearance at an observation
point, but at the instant of the ST passage in the magneto-conjugate
region. This implies that the TWP formation depends on a
number of phenomena at magneto-conjugate points and in the
magnetic field line joining these points.

Seasonal dependence and connection with processes in the
magneto-conjugate point indicate that MSTWPs have electrodynamic
nature. Data from simultaneous optical observations of
periodic structures in Japan and Australia also confirm this
connection. In recent years, geomagnetic conjugate observations of
TIDs using 630-nm airglow imagers were reported. Otsuka et
al. (2004) first reported a symmetric pattern of nighttime MSTIDs
using 630-nm airglow images obtained at geomagnetic conjugate
points in the northern and southern hemispheres. MSTIDs are often
observed at the middle latitudes over Japan at night, regardless of
geomagnetic activity. These nighttime MSTIDs generally have
wavelengths of a few hundred kilometers and propagate
southwestward at 50-100 m/s [Shiokawa et al., 2005]. Shiokawa et
al. (2005) concluded that the symmetric structures of MSTIDs
indicated strong electrodynamic coupling between the two
hemispheres through the geomagnetic field line. However it should
be noted that all above mentioned papers did not consider the
mechanism of generation of studied MSTIDs as a consequence of the
movement of the solar terminator.

Ionospheric processes in magneto-conjugate regions have long
been studied [Krinberg and Tashchilin, 1984]. However, these
investigations did not concern wave processes. The exception was
work by Abramchuk et al. (1987). The authors ascertained that the
probability of occurrence of the mid-latitude layer Es increased
when the terminator passed through the magneto-conjugate region.
They advanced the hypothesis allowing the observed phenomenon to
be explained by the Alfven wave propagation along the magnetic
field line. But no experimental evidence was given for the
recording of corresponding wave process.

Using a new law to the mid-latitude model Sami2 of the ionosphere,
Huba et al. (2000) found that ion sound waves can be generated in
the topside low-latitude ionosphere at sunrise and sunset. The
waves can persist $\approx$ 1 - 3 h at altitudes above the O+/H+
transition altitude ($\approx$ 1000 kin) with periods about 10
min. The waves result from the rapid heating and cooling of
the lower ionosphere that occurs at sunrise and sunset. At
sunrise, photoelectron heating produces strong upward plasma flows
along the geomagnetic field. These flows lead to a local
compression and heating of the plasma at the apex of the field
line, which in turn generates ion sound waves. At sunset, the waves
are produced by a rapid cooling of the plasma. According to [Huba
et al., 2006], the velocity of wave propagation is about 6 km/s,
while the wavelength is close to the length of the corresponding
magnetic line.

Our study is the first experimental proof for the correctness of the
model [Huba et al., 2000]. Actually, we found out a NEW
PHENOMENON and provided the experimental proof for a possible
detection of ST-generated ion acoustic waves (oscillation periods
of about 20 minutes) using modern methods of ionospheric
diagnostics. As such oscillations are sometimes connected with
the terminator passage in the ionospheric region conjugated via
the geomagnetic field, their transportation by some magnetospheric
MHD waves would appear reasonable. Alfven or slow
magneto sonic (SMS) waves propagate along the magnetic field. The
periods of the observed oscillations are far beyond the minimum
periods of the proper Alfven waves at these latitudes ($\approx$10
s), but correspond to the periods of the first harmonics of
stationary SMS waves propagating along geomagnetic field lines
[Leonovich et al., 2006]. In that study the authors conclude that
the ionosphere plays no part in either generation or SMS wave
absorption. This conclusion stems from the fact that all
electromagnetic components and transverse components of plasma
oscillations tend to zero in the ionosphere. But the said study
also shows that the longitudinal component of the plasma
oscillation velocity does not vanish in the ionosphere (also see
[Cheremnykh and Parnovsky, 2004]. It is just this oscillation
component that is responsible for the electron density modulation
that is detected by TEC recording.

Skewness of seasonal dependence in Japan is noteworthy (Fig. 3).
In summer, wave packets appear before the sunset, but they are not
registered at the winter magneto-conjugate area. This implies that
the plasma modulated stream moves in the only one direction - from
the winter area after sunset to the summer daytime
magneto-conjugate area.

\section{Conclusion}
\label{SPE-sect-7} Main results of this study may be
summarized as follows. We obtained the first experimental evidence
for the magnetohydrodynamic (MHD) nature of ionospheric
medium-scale travelling wave packets (MSTWP). Data were used from
total electron content (TEC) measurements made at the dense Japanese
network GPS/GEONET (1220 stations) in 2008-2009. The
diurnal, seasonal and spectral MSTWP characteristics were found to be specified
by the solar terminator (ST) dynamics. MSTWPs are the chain of
narrow-band TEC oscillations with single packet's duration of
about 1-2 hours and oscillation periods of 10-20 minutes. Its
total duration is about 4--6 hours. The MSTWP spatial structure is
characterized by the high degree of anisotropy and coherence at
the distance of more than 10 wavelengths. MSTWP's direction of
travelling is characterized by the hight directivity independently
of seasons. The daytime MSTWPs occurrence rate is high in
winter and during equinoxes. Most of the daytime MSWPs propagate
with the velocity of 130$\pm$52 m/s southeastward (155$\pm$28${}^\circ$).
The nighttime MSTIDs occurrence rate peaks in summer. They
propagate at
110$\pm$43 m/s southwestward (245$\pm$15${}^\circ$). These features are consistent with previous MSTID
statistics from 630-nm airglow observations in Japan. In
winter, MSTWPs in the northern hemisphere are observed 3-4 hours
after the morning ST passage. In summer, MSTWPs are detected 1.5-2
hours before the evening ST occurrence at the observation point,
 but at the moment of the evening ST passage in the
magneto-conjugate area. Both the high Q-factor of oscillatory
system and synchronization of MSTWP occurrence with the solar
terminator passage at the observation point and in the magneto-conjugate
 area testify the MHD nature of ST-excited MSTWP
generation. The obtained results are the first experimental
evidence for the hypothesis of the ST-generated ion sound waves
[Huba et al, 2000].

There are many existing and supposed sources of MSAGWs, which
make a random interference field of the neutral atmosphere wave
disturbance. The result is a random MS intensity distribution of
electron density in the ionosphere with chaotic change of the
apparent travel direction. This can be recorded some hours before
the passage of the morning and evening ST over the observation
point or in the magneto-conjugate area. When ST arrives, a
strongly regular structure of wave disturbance of MHD origin can
be observed. This structure overlaps the random interference field
of different sources [Hocke and Schlegel, 1996].

We are aware that this study has revealed only the key averaged
patterns of this phenomena, and we hope that it would give impetus
to more detailed investigations. However, gaining a
better understanding of the formation mechanism for MSTWP
disturbances requires modelling and further experimental investigations of
the whole process.

Our results are important for the development of ionosphere irregularity
physics and modelling of the transionosphere radio wave propagation.
The obtained results may be used for development of MSTIDs model
necessary for different applications. We should know the time of
occurrence of such structures and the direction of wave disturbance
front elongation to optimize performance of different satellite
radio technical systems. This knowledge is also important for
finding possible earthquakes precursors in the given range of TEC oscillation
periods, for detecting ionospheric response to hurricanes and
tornadoes, etc. Finding the said disturbances that have seismic
and meteorological origin is impossible because of the presence of the ST-generated
regular wave structure(up to one third of a day).

Spatial-temporal properties and modelling of MSTWP are the aim of
our future works.

A new era for studying terminator waves and medium-scale
ionospheric disturbances begins: this provides a better space-time
resolution, global observations, and a magnetohydrodynamic
conception of nature of these disturbances…
%

\section*{Acknowledgments}
The authors thank Profs. A.S. Leonovich, V.A. Mazur and V.M.
Somsikov for their interest to the work and fruitful discussion,
and Dr. Zhivetiev I.V., Dr. Ding Feng and Prof. S.G.
Jin for their assistance in data receiving.
We acknowledge the Scripps Orbit and Permanent Array Center
(SOPAC), the Crustal Dynamics Data Information System (CDDIS) and
GEONET for providing GPS data used in this study. The work was
supported by the Interdisciplinary integral project of SB RAS N
56, the RFBR-GFEN grant N 06-05-39026 and RFBR grant 07-05-00127.


\end{document}